\documentclass[a4paper,10pt]{article}
\usepackage{amsmath}
\usepackage{amssymb}

\newlength{\minitwocolumn}
\makeatletter
\@addtoreset{equation}{section}
\makeatother

\title{\bf Difference equations for 
the higher rank XXZ model with a boundary}

\author{Takeo Kojima$^*$ and 
Yas-Hiro Quano$^\dagger$}

\date{\it ${}^*$Department of Mathematics,
College of Science and Technology, 
Nihon University, Chiyoda-ku, Tokyo 101-0062, Japan \\
\it ${}^\dagger$Department of Medical Electronics, 
Suzuka University of Medical Science \\
      \it Kishioka-cho, Suzuka 510-0293, Japan}
\begin{document}

\maketitle
\begin{abstract}
The higher rank analogue of the XXZ model 
with a boundary is considered on the basis of 
the vertex operator approach. 
We derive difference equations of 
the quantum Knizhnik-Zamolodchikov type for 
$2N$-point correlations of the model. 
We present infinite product formulae
of two point functions with free boundary condition
by solving those difference equations with $N=1$.
\end{abstract}

\section{Introduction}
Representation theory of the affine quantum group
plays an important role in the description 
of solvable lattice models and massive 
integrable quantum field theories in 
two dimentions \cite{FR,Sm1,JM}. 
For models with the affine quantum group symmetry 
the difference analogue of the 
Knizhnik-Zamolodchikov equations 
(quantum Knizhnik-Zamolodchikov equations) 
are satisfied by both correlation functions
and form factors \cite{FR,S,JMN}.

Integrable models with boundary reflection 
have been also studied in lattice models and 
massive quantum theories. The boundary interaction 
is specfied by the reflection matrix $K$ 
for lattice models \cite{Skl}, and by the boundary 
$S$-matrix for massive quantum theories \cite{GZ}.
It is shown in \cite{JKKKM} that
the space of states of the boundary XXZ model 
can be described in terms of
vertex operators associated with the bulk 
XXZ model \cite{JM}.
The explicit bosonic formulae of the boundary
vacuum of the boundary XXZ model were obtained 
by using the bosonization of the vertex operators 
\cite{JKKKM}. This approach is also relevant for 
other various models \cite{MW, HSWY, FK, FKQ, H}.

It is shown in \cite{JKKMW} that correlation functions 
and form factors in semi-infinite XXZ/XYZ spin chains 
with integrable boundary conditions satisfy the boundary 
analogue of the quantum Knizhnik-Zamolodchikov equation. 
In this paper we establish 
the similar results for the $U_q(\widehat{sl_n})$
-analogue of 
XXZ spin chain with a boundary magnetic field $h$:
\begin{eqnarray}
{\cal H}_B=\sum_{k=1}^\infty
\left\{q\sum_{a,b=0 \atop{a>b}}^{n-1}
e_{aa}^{(k+1)}e_{bb}^{(k)}+q^{-1}
\sum_{a,b=0 \atop{a<b}}^{n-1}e_{aa}^{(k+1)}
e_{bb}^{(k)}
-\sum_{a,b=0 \atop{a \neq b}}^{n-1}
e_{ab}^{(k+1)}e_{ba}^{(k)}\right\}
\label{HamiltonianB}\\
+\frac{1-q^2}{2q} \left\{
\sum_{a=0}^{L-1}e_{aa}^{(1)}
-\sum_{a=M}^{n-1}e_{aa}^{(1)}\right\}
+h \sum_{a=L}^{M-1}e_{aa}^{(1)},\nonumber
\end{eqnarray}
where $-1<q<0$ and
$0\leqq L \leqq M \leqq n-1$.
On the basis of the boundary vacuum states 
constructed in \cite{FK} we derive 
the boundary analogue of the quantum 
Knizhnik-Zamolodchikov equations for 
the correlation functions in 
the higher rank XXZ model with
a boundary. We also obtain the 
two point functions by solving the simplest difference 
equations for free boundary condition. 

The rest of this paper is organized as follows.
In section 2 we review the vertex operator approach
for the higher rank XXZ model with a boundary.
In section 3 we derive the boundary 
quantum Knizhnik-Zamolodchikov equations for 
the $2N$-point correlation functions.
In section 4 we obtain the two point functions
by solving the difference 
equation with $N=1$ for free boundary condition. 
In Appendix A we summarize the results of 
the bosonization of the vertex operators 
in $U_q(\widehat{sl_n})$ \cite{Koy}.
In Appendix B we summarize the bosonic formulae
of the boundary vacuum states \cite{FK}.

\section{Formulation}
The higher rank XXZ model with boundary reflection 
was formulated in \cite{FK} in terms of the vertex 
operators of the quantum affine 
group $U_q(\widehat{sl_n})$.
For readers' convenience let us briefly review
the results in \cite{FK}. 

Throughout this paper we fix 
$n \in {\mathbb{N}_{\geqq 2}}$, and also fix 
$q$ such that $-1<q<0$. 
The model is labeled by the three parameters 
$i, L, M$ such that 
$0\leqq L \leqq M \leqq n-1$ and $i\in \{ L, M \}$. 
In this paper we consider the following
three cases:
$$
\begin{array}{rl}
(C1) & 0\leqq L=M=i\leqq n-1, \\
(C2) & 0\leqq L=i <M\leqq n-1, \\
(C3) & 0\leqq L<M=i\leqq n-1. 
\end{array}
$$
In what follows we denote 
the $q$-integer $(q^k-q^{-k})/(q-q^{-1})$ by $[k]$, 
and we use the following simbols:
\begin{eqnarray}
b(z)=\frac{q-q^{-1}z}{1-z},~~~
c(z)=\frac{q-q^{-1}}{1-z}.
\end{eqnarray}
The nonzero entries of the R-matrix
$R^{(i)VV}(z)$ are given by
\begin{eqnarray}
R^{(i)VV}(z)_{j_1,j_2}^{k_1,k_2}=r^{(i)VV}(z)\times
\left\{\begin{array}{cc}
1,&~~j_1=j_2=k_1=k_2\\
b(q^2z),&~~j_1=k_1\neq j_2=k_2,\\
-qc(q^2z),&~~j_1=k_2 < j_2=k_1,\\
-qz c(q^2z),&~~j_1=k_2 > j_2=k_1. 
\end{array}
\right.\label{RVV}
\end{eqnarray}
Here the scalar functions are 
\begin{eqnarray}
r^{(i)VV}(z)=z^{-\delta_{i,0}}
\frac{(q^2z^{-1};q^{2n})_\infty 
(q^{2n}z;q^{2n})_\infty}
{(q^2z;q^{2n})_\infty 
(q^{2n}z^{-1};q^{2n})_\infty}, 
\end{eqnarray}
where 
$$
(z;p_1 , \cdots , p_m )_\infty
=\prod_{k_1, \cdots , k_m =0}^\infty 
(1-zp_1^{k_1}\cdots p_m^{k_m}). 
$$
The boundary K-matrix $K^{(i)}(z)$ is a diagonal matrix,
whose diagonal elements are given by
\begin{eqnarray}
K^{(i)}(z)_j^j=\frac{\varphi^{(i)}(z)}
{\varphi^{(i)}(1/z)}
\times
\left\{\begin{array}{cc}
z^2,& 0\leqq j \leqq L-1,\\
\frac{\displaystyle 1-rz}
{\displaystyle 1-r/z},& L\leqq j \leqq M-1,\\
1,& M\leqq j \leqq n-1,
\end{array}\right.\label{K}
\end{eqnarray}
where we have set
\begin{eqnarray}
\varphi^{(i)}(z)=z^{\delta_{i,0}-1}
\frac{(q^{2n+2}z^2;q^{4n})_\infty}{(q^{4n}z^2;q^{4n})_\infty}
\times\left\{\begin{array}{cc}
1,&~{\rm for}~(C1),\\
\frac{ \displaystyle
(rq^{2n}z;q^{2n})_\infty}
{\displaystyle
(rq^{2n-2M+2L}z;q^{2n})_\infty},&~
{\rm for}~ (C2),\\
\frac{\displaystyle
(r^{-1}z;q^{2n})_\infty}{
\displaystyle
(r^{-1}q^{2M-2L}z;q^{2n})_\infty},&~
{\rm for}~(C3).
\end{array}\right.
\end{eqnarray}
They satisfy the boundary Yang-Baxter equation: 
\begin{eqnarray}
K_2^{(i)}(z_2)R_{21}^{(i)}(z_1z_2)
K_1^{(i)}(z_1)R_{12}^{(i)}(z_1/z_2)=
R_{21}^{(i)}(z_1/z_2)K_1^{(i)}(z_1)R_{12}
^{(i)}(z_1z_2)K_2^{(i)}(z_2).
\label{BYBE}
\end{eqnarray}

Let $V={\mathbb C}v_0
\oplus \cdots \oplus {\mathbb C}v_{n-1}$ 
be the basic representation of $U_q(sl_n)$,
and let $V_z$ be the evaluation representation
of $U_q(\widehat{sl_n})$ in the homogeneous 
picture. 
Let $V(\Lambda_i)$ be the irreducible
highest weight module with the
level 1 highest weight $\Lambda_i~(i=0,\cdots,n-1)$.
The type-I vertex operator 
$\Phi^{(i,i+1)}(z)$ 
is an intertwining operator of $U_q(\widehat{sl_n})$
defined by
\begin{eqnarray}
\Phi^{(i,i+1)}(z)
:V(\Lambda_{i+1})\to V(\Lambda_i)
\hat{\otimes} V_z, 
\end{eqnarray}
where the superscripts $i, i+1$ should be 
interpreted as elements in $\mathbb{Z}_n$. 
Let us define the component of the vertex operators
$\Phi^{(i,i+1)}_j(z)$ as follows.
\begin{eqnarray}
\Phi^{(i,i+1)}(z)|u\rangle=
\sum_{j=0}^{n-1}
\Phi^{(i,i+1)}_j(z)|u\rangle \otimes v_j,
~{\rm for}~|u\rangle \in V(\Lambda_{i+1}).
\end{eqnarray}
The dual type-I vertex operator 
$\Phi^{*(i+1,i)}(z)$ is an intertwining
operator of $U_q(\widehat{sl_n})$ defined by
\begin{eqnarray}
\Phi^{*(i+1,i)}(z):V(\Lambda_{i})\otimes V_z
\to \hat{V}(\Lambda_{i+1}).
\end{eqnarray}
Let us define the components of the dual vertex 
operators $\Phi_j^{*(i+1,i)}(z)$ as follows.
\begin{eqnarray}
\Phi^{*(i+1,i)}(z)(|u\rangle \otimes v_j)
=\Phi^{*(i+1,i)}_j(z)|u\rangle,~
~{\rm for}~|u\rangle \in V(\Lambda_{i}).
\end{eqnarray}

Let us summarize here the properties of 
the vertex operators: 

\noindent{\bf Commutation relations} 
The vertex operators satisfy the following commutation
relation: 
\begin{eqnarray}
\Phi^{(i-2,i-1)}_{j_2}(z_2)
\Phi^{(i-1,i)}_{j_1}(z_1)=
\sum_{j_1', j_2'=0
\atop{j_1'+j_2'=j_1+j_2}}^{n-1}
R^{(i) V V }(z_1/z_2)_{j_1,j_2}^{j_1',j_2'}
\Phi^{(i-2,i-1)}_{j_1'}(z_1)\Phi^{(i-1,i)}_{j_2'}(z_2). 
\label{commuteVV}
\end{eqnarray}
Concerning other commutation relations (\ref{commuteV*V}), 
(\ref{commuteVV*}) and (\ref{commuteV*V*}), see section 3. 

\noindent{\bf Normalizations} 
We adopt the following normalizations: 
\begin{eqnarray}
\Phi^{(i,i+1)}(z)|\Lambda_{i+1}\rangle=
|\Lambda_i\rangle \otimes v_i+\cdots,~~
\Phi^{*(i+1,i)}(z)|\Lambda_i\rangle
\otimes v_i=|\Lambda_{i+1}\rangle +\cdots,
\end{eqnarray}
where $|\Lambda_i\rangle$ is the highest weight
vector of $V(\Lambda_i)$.

\noindent{\bf Invertibility} 
They satisfy the following inversion relation:
\begin{eqnarray}
g_n \Phi_j^{(i-1,i)}(z)\Phi_j^{*(i,i-1)}(z)={\rm id}, 
\label{eq:inv-rel}
\end{eqnarray}
where 
$$
g_n=\frac{(q^2;q^{2n})_\infty}
{(q^{2n};q^{2n})_\infty}.
$$

~

We define the normarized transfer matrix by
\begin{eqnarray}
T_B^{(i)}(z)=g_n
\sum_{j=0}^{n-1}\Phi_j^{*(i,i-1)}(z^{-1})
K^{(i)}(z)_j^j\Phi_j^{(i-1,i)}(z),~~
\label{transfer}
\end{eqnarray}
Let the space ${\cal H}^{(i)}$ be the span of
vectors $|p\rangle=\otimes_{k=1}^\infty v_{p(k)}$,
where $p: {\mathbb{N}} \to \mathbb{Z}/n \mathbb{Z}$
satisfies the asymptotic condition
\begin{eqnarray}
p(k)=k+i \in \mathbb{Z}/n \mathbb{Z},~~
{\rm for}~k \gg 1.
\end{eqnarray}
As usual, the transfer matrix (\ref{transfer}) and the 
Hamiltonian (\ref{HamiltonianB})
are related by
\begin{eqnarray}
\left.\frac{d}{dz}T_B^{(i)}(z)\right|_{z=1}=
\frac{2q}{1-q^2}{\cal H}_B+const,~~
{\rm for}~~~h=\frac{r+1}{r-1}\times \frac{1-q^2}{2q}.
\label{HT}
\end{eqnarray}
Note that the left hand side act on the space $V(\Lambda_i)$ 
while the right hand side acts on the space 
${\cal H}^{(i)}$. Thus 
we can make the following identification: 
\begin{eqnarray}
V(\Lambda_i) \simeq {\cal H}^{(i)}.
\end{eqnarray}
The boundary ground state and the dual boundary 
ground state are characterized by
\begin{eqnarray}
T_B^{(i)}(z)|i\rangle_B=|i\rangle_B,~(i=0,\cdots,n-1),
\label{eigen-1}\end{eqnarray}
and
\begin{eqnarray}
~_B\langle i |T_B^{(i)}(z)=~_B\langle i |,~~(i=
0,\cdots, n-1).\label{eigen-2}
\end{eqnarray}
Using the inversion relation,
the eigenvalue problems (\ref{eigen-1}) 
and (\ref{eigen-2}) are reduced to
\begin{eqnarray}
K^{(i)}(z)_j^j\Phi_j^{(i-1,i)}(z)|i\rangle_B=
\Phi_j^{(i-1,i)}(z^{-1})|i\rangle_B,
\label{eq:KP}
\end{eqnarray}
and
\begin{eqnarray}
K^{(i)}(z)_j^j~_B\langle i|\Phi_j^{*(i,i-1)}(z^{-1})=
~_B\langle i |\Phi_j^{*(i,i-1)}(z).
\label{eq:KP*}
\end{eqnarray}
The bosonizations of vertex operators are given in 
\cite{Koy}.
The bosonic formulae of the boundary vacuum are 
given in \cite{FK}.
For readers' convenience we summarize the 
bosonizations of vertex operators
in Appendix A
and the bosonic formula of the boundary vacuum in Apendix B.

\section{Boundary quantum Knizhnik-Zamolodchikov equations}
The purpose of this section is to derive
the $q$-difference equations for 
the correlation function
of the higher rank XXZ spin chain with a boundary
magnetic field.
For $U_q(\widehat{sl_2})$
case \cite{JKKMW}, the said difference equations are 
based on the duality relation of vertex operators
$$
\Phi^*_\epsilon(\zeta)=\Phi_{-\epsilon}(-q^{-1}\zeta), 
$$
in addition to (\ref{commuteVV}), (\ref{eq:KP}) 
and (\ref{eq:KP*}). 
For $n>2$ case, however, the dual vertex operator
$\Phi_j^*(z)$ is written in terms of 
$(n-1)$-st determinant of $\Phi_j(z)$'s.
Thus it is not convenient
to use the duality relation for the present case.

For $n>2$, 
we use the explicit formulae of the boundary states
to derive the boundary quantum Knizhnik-Zamolodchikov
equations. 
In this section we establish 
the following simple relations: 
\begin{equation}
\begin{array}{rcl}
\Phi^{*(i+1,i)}_j(q^nz)|i\rangle_B&=&
K^{*(i)}(z)_j^j\Phi^{*(i+1,i)}_j(q^n/z)|i\rangle_B,~
(j=0,\cdots,n-1),\\
~_B\langle i |\Phi^{(i,i+1)}_j(1/(q^nz))&=&
{K}^{*(i)}(z)_j^j~_B\langle i |
\Phi^{(i,i+1)}_j(z/q^n),~
(j=0,\cdots,n-1),
\end{array}
\label{eq:dual-KP}
\end{equation}
where the functions $K^{(i)*}(z)_j^j$ are given by
(\ref{C1}), (\ref{C2}) and (\ref{C3}). 
The relations (\ref{eq:dual-KP}) in addition to 
the commutation relations (\ref{commuteVV}), 
(\ref{commuteV*V}), (\ref{commuteVV*}) and 
(\ref{commuteV*V*}) 
imply the $q$-difference equations of the present model.

\subsection{Boundary state}
In this subsection we use the symbols 
$P^*(z), Q^*(z), R^-(z), S^-(z)$, which are 
bosons defined in Appendix A. See Appendix A 
as for the definitions. 

Let us first consider
consider $``(C1)~0\leqq L=M=i\leqq n-1''$-case.
Let us show the following relation: 
\begin{eqnarray}
\Phi^{*(i+1,i)}_j(q^nz)|i\rangle_B=
K^{*(i)}(z)_j^j\Phi^{*(i+1,i)}_j(q^n/z)|i\rangle_B,~
(j=0,\cdots,n-1), 
\label{eqn:Master}
\end{eqnarray}
where 
\begin{eqnarray}
K^{*(i)}(z)_j^j=
\frac{\varphi^{*(i)}(z)}{\varphi^{*(i)}(1/z)}
\times
\left\{
\begin{array}{cl}
z^2,&(0\leqq j \leqq L-1=i-1),\\
1,&(
i=L \leqq j \leqq n-1),
\end{array}
\right.~~{\rm for}~~(C1),\label{C1}
\end{eqnarray}
and 
\begin{eqnarray}
\varphi^{*(i)}(z)=z^{\delta_{i,0}}
\frac{(q^{4n}z^2;q^{4n})_\infty }
{(q^{2n+2}z^2;q^{4n})_\infty }.
\end{eqnarray}
Multiply the both sides of (\ref{eqn:Master}) 
by $k^{(i)}_j(z)\varphi^{*(i)}(1/z)$, where 
\begin{eqnarray}
k^{(i)}_j(z)=\left\{\begin{array}{cc}
z^{-1},& 0\leqq j \leqq i-1,\\
1,& i\leqq j \leqq n-1.
\end{array}\right.
\end{eqnarray}
Then 
the RHS of (\ref{eqn:Master}) is obtained 
from the LHS by changing $z\rightarrow 1/z$. 

Bosonization formulae of $P^*(z), Q^*(z)$ and 
$|i\rangle_B$ imply the identity 
\begin{eqnarray}
e^{Q^*(q^nz)}|i\rangle_B=
\frac{(q^{2n+2}z^{-2};q^{4n})_\infty}
{(q^{4n}z^{-2};q^{4n})_\infty}e^{P^*(q^n/z)}|i\rangle_B. 
\end{eqnarray}
By using this identity we have 
\begin{equation}
k^i_0(z)\varphi^{*(i)}(1/z)\Phi^{*(i+1,i)}_0(q^nz)
|i\rangle_B
=c^*_0 
e^{P^*(q^nz)+P^*(q^n/z)}e^{\bar{\Lambda}_1}|i\rangle_B, 
\label{eq:P_0}
\end{equation}
where $c_0^*$ is some constant. 
The relation (\ref{eqn:Master}) with $j=0$ 
follows form the fact that RHS of (\ref{eq:P_0}) is 
symmetric under $z\rightarrow 1/z$. 

Invoking the bosonization
of the dual vertex operators, 
we also have for $j>0$ as follows: 
\begin{equation}
\begin{array}{cl}
& k^{(i)}_j(z)\varphi^{*(i)}(1/z)\Phi^{*(i+1,i)}_j (q^nz) 
|i\rangle_B \\
=&\displaystyle c_j^* \oint \frac{dw_1}{w_1}
\cdots \oint \frac{dw_j}{w_j} k^{(i)}_j (z)
{\rm Int}(z,w_1,w_2,\cdots,w_{j})
e^{P^*(q^nz)+P^*(q^n/z)}\\
\times&\displaystyle e^{R_1^-(q^{n+1}w_1)
+R_1^-(q^{n+1}/w_1)+\cdots
+R_j^-(q^{n+1}w_j)+
R_j^-(q^{n+1}/w_j)
}e^{\bar{\Lambda}_1}|i\rangle_B, 
\end{array}
\end{equation}
where $c_j^*$'s are some constants. 
Here we set the integrand: 
\begin{eqnarray}
{\rm Int}(w_0,w_1,\cdots,w_{j})
=\frac{w_j \prod_{k=1}^{j}\left\{(1-w_k^{-2})
w_k^{-\delta_{k,i}}(1-qw_{k-1}w_{k})\right\}
}{\prod_{k=1}^{j} D(w_{k-1},w_{k})},
\end{eqnarray}
where
$$
D(w_1,w_2)=
(1-qw_1w_2)(1-qw_1/w_2)(1-qw_2/w_1)(1-q/(w_1w_2)).
$$
Thus the relation (\ref{eqn:Master}) with $j>0$ 
follows from the identities 
\begin{eqnarray}
\sum_{\epsilon_1=\pm,\cdots,\epsilon_j=\pm}
\left\{k^{(i)}_j(z^{-1})
{\rm Int}(z,w_1^{\epsilon_1},\cdots,
w_{j}^{\epsilon_j})-k^{(i)}_j(z)
{\rm Int}(z^{-1},w_1^{\epsilon_1},\cdots,w_{j}^{\epsilon_j})
\right\}=0. 
\end{eqnarray}

Let us consider $``(C2)~0\leqq L=i < M\leqq n-1''$-case.
From the same arguments as for $(C1)$, we have
\begin{eqnarray}
K^{*(i)}(z)_j^j=
\frac{\varphi^{*(i)}(z)}{\varphi^{*(i)}(1/z)}
\times
\left\{
\begin{array}{cc}
z^2,&(0\leqq j \leqq L-1),\\
\frac{\displaystyle 1-q^{n-2M+2L}rz}
{\displaystyle 1-q^{n-2M+2L}rz^{-1}},&(L\leqq j \leqq M-1),\\
1,&(M\leqq j \leqq n-1),
\end{array}
\right.~~{\rm for}~~(C2),\label{C2}
\end{eqnarray}
where we have set
\begin{eqnarray}
\varphi^{*(i)}(z)=z^{\delta_{i,0}}
\frac{(q^{4n}z^2;q^{4n})_\infty (rq^{n}z;q^{2n})_\infty}
{(q^{2n+2}z^2;q^{4n})_\infty (rq^{n+2L-2M}z;q^{2n})_\infty}.
\end{eqnarray}
In this case the following relations are useful: 
\begin{eqnarray}
e^{Q^*(q^nz)}|0\rangle_B=
\frac{(rq^{3n-2M}z^{-1};q^{2n})_\infty
(q^{2n+2}z^{-2};q^{4n})_\infty}
{(rq^nz^{-1};q^{2n})_\infty (q^{4n}z^{-2};q^{4n})_\infty}
e^{P^*(q^n/z)}|0\rangle_B,\\
e^{Q^*(q^nz)}|i\rangle_B=
\frac{(rq^{n+2L-2M}z^{-1};q^{2n})_\infty
(q^{2n+2}z^{-2};q^{4n})_\infty}
{(rq^nz^{-1};q^{2n})_\infty (q^{4n}z^{-2};q^{4n})_\infty}
e^{P^*(q^n/z)}|i\rangle_B,~(i\geqq 1),
\end{eqnarray}
and
\begin{eqnarray}
e^{S_j^-(w)}|i\rangle_B=
g_j^{(i)}(w)
e^{R_j^-(q^{2(n+1)}/w)}|i\rangle_B,
\end{eqnarray}
where
\begin{eqnarray}
g_j^{(0)}(q^{n+1}w)=
\left\{\begin{array}{cc}
(1-1/w^{2})(1-q^{-n+2M-L}/(rw)),&~j=L,\\
(1-1/w^{2})(1-q^{n-M}r/w),&~j=M,\\
(1-1/w^{2}),&~j\neq L,M,
\end{array}\right.
\end{eqnarray}
and
\begin{eqnarray}
g_j^{(i)}(q^{n+1}w)=
\left\{\begin{array}{cc}
\frac{\displaystyle (1-1/w^{2})}
{\displaystyle (1-rq^{n-2M+L}/w)},&~j=L,\\
(1-1/w^{2})(1-q^{n-M}r/w),&~j=M,\\
(1-1/w^{2}),&~j\neq L,M, 
\end{array}\right.~~(i\geqq 1).
\end{eqnarray}

Let us consider $``(C3)~0\leqq L< M=i\leqq n-1''$-case.
Repeating the same procedure as in $(C1)$, we have
\begin{eqnarray}
K^{*(i)}(z)_j^j=
\frac{\varphi^{*(i)}(z)}{\varphi^{*(i)}(1/z)}
\times
\left\{
\begin{array}{cc}
z^2,&(0\leqq j \leqq L-1),\\
\frac{\displaystyle 1-q^{n+2M-2L}r^{-1}z}
{\displaystyle 1-q^{n+2M-2L}r^{-1}z^{-1}},&(L\leqq j \leqq M-1),\\
1,&(M\leqq j \leqq n-1),
\end{array}
\right.~~{\rm for}~~(C3), \label{C3}
\end{eqnarray}
where we set
\begin{eqnarray}
\varphi^{*(i)}(z)=
\frac{(q^{4n}z^2;q^{4n})_\infty (r^{-1}q^{n}z;q^{2n})_\infty}
{(q^{2n+2}z^2;q^{4n})_\infty (r^{-1}q^{n+2M-2L}z;q^{2n})_\infty}.
\end{eqnarray}
In this case the following relations are useful: 
\begin{eqnarray}
e^{Q^*(q^nz)}|i\rangle_B=
\frac{(r^{-1}q^{2M-n}z^{-1};q^{2n})_\infty 
(q^{2n+2}z^{-2};q^{4n})_\infty}
{(r^{-1}q^n z^{-1};q^{2n})_\infty 
(q^{4n}z^{-2};q^{4n})_\infty}e^{P^*(q^n/z)}|i\rangle_B,~(L=0),\\
e^{Q^*(q^nz)}|i\rangle_B=
\frac{(r^{-1}q^{2M-2L+n}z^{-1};q^{2n})_\infty 
(q^{2n+2}z^{-2};q^{4n})_\infty}
{(r^{-1}q^n z^{-1};q^{2n})_\infty 
(q^{4n}z^{-2};q^{4n})_\infty}
e^{P^*(q^n/z)}|i\rangle_B,~(L\geqq 1), 
\end{eqnarray}
and
\begin{eqnarray}
e^{S_j^-(w)}|i\rangle_B=
g_j^{(i)}(w)
e^{R_j^-(q^{2(n+1)}/w)}|i\rangle_B,
\end{eqnarray}
where
\begin{eqnarray}
g_j^{(i)}(q^{n+1}w)=
\left\{\begin{array}{cc}
(1-1/w^{2})(1-q^{-n+2M-L}/(rw)),&~j=L,\\
\frac{\displaystyle (1-1/w^{2})}
{\displaystyle (1-q^{M-n}/(rw))},&~j=M,\\
(1-1/w^{2}),&~j\neq L,M,
\end{array}\right.
\end{eqnarray}

\subsection{Dual boundary state}
From the same arguments as for the boundary state case,
we can show the following relation: 
\begin{eqnarray}
~_B\langle i|\Phi^{(i,i+1)}_j(1/(q^nz))=
{K}^{*(i)}(z)_j^j
~_B\langle i|\Phi^{(i,i+1)}_j(z/q^n).
\end{eqnarray}
For each case the following relations are useful: \\
$(C1)$-case: $0\leqq L=M=i \leqq n-1$
\begin{equation}
\begin{array}{rcl}
~_B\langle i |e^{P(z/q^n)}&=&
\dfrac{(q^{2n+2}z^2;q^{4n})_\infty}
{(q^{4n}z^2;q^{4n})_\infty}~_B\langle i|e^{Q(1/(q^nz))},
\\[3mm]
~_B\langle i|e^{S_j^-(w)}&=&g_j^{*(i)}(w)
~_B\langle i|e^{R_j^-(q^2/w)},
\end{array}
\end{equation}
where
\begin{eqnarray}
g_j^{*(i)}(qw)=(1-w^2).
\end{eqnarray}
$(C2)$-case: $0\leqq L=i <M \leqq n-1$
\begin{equation}
\begin{array}{rcl}
~_B\langle i |e^{P(z/q^n)}&=&
\dfrac{(q^{2n+2}z^2;q^{4n})_\infty
(rq^{n+2L-2M}z;q^{2n})_\infty}
{(q^{4n}z^2;q^{4n})_\infty
(rq^nz;q^{2n})_\infty}~_B\langle i|e^{Q(1/(q^nz))}, 
\\[3mm]
~_B\langle i|e^{S_j^-(w)}&=&g_j^{*(i)}(w)
~_B\langle i|e^{R_j^-(q^2/w)},
\end{array}
\end{equation}
where
\begin{eqnarray}
g_j^{*(0)}(qw)=
\left\{
\begin{array}{cc}
\frac{\displaystyle(1-w^2)}{
\displaystyle (1-q^{-L}w/r)},&j=L,\\
\frac{\displaystyle (1-w^2)}{
\displaystyle (1-q^{2L-M}rw)},&j=M,\\
(1-w^2),&j\neq L,M,
\end{array}\right.
\end{eqnarray}
and
\begin{eqnarray}
g_j^{*(i)}(qw)=
\left\{
\begin{array}{cc}
(1-w^2)
(1-q^{L}rw),&j=L,\\
\frac{\displaystyle (1-w^2)}{
\displaystyle (1-q^{2L-M}rw)},&j=M,\\
(1-w^2),&j\neq L,M,
\end{array}\right.~(i\geqq 1).
\end{eqnarray}
$(C3)$-case: $0\leqq L<M=i \leqq n-1$
\begin{equation}
\begin{array}{rcl}
~_B\langle i |e^{P(z/q^n)}&=&
\dfrac{(q^{2n+2}z^2;q^{4n})_\infty
(q^{n+2M-2L}r^{-1}z;q^{2n})_\infty}
{(q^{4n}z^2;q^{4n})_\infty
(q^{n}r^{-1}z;q^{2n})_\infty}~_B\langle i|e^{Q(1/(q^nz))} 
\\[3mm]
~_B\langle i|e^{S_j^-(w)}&=&g_j^{*(i)}(w)
~_B\langle i|e^{R_j^-(q^2/w)},
\end{array}
\end{equation}
where
\begin{eqnarray}
g_j^{*(i)}(qw)=
\left\{
\begin{array}{cc}
\frac{\displaystyle (1-w^2)}{
\displaystyle (1-q^{-L}w/r)},&j=L,\\
(1-w^2)(1-q^{M-2L}w/r),&j=M,\\
(1-w^2),&j\neq L,M,
\end{array}\right.
\end{eqnarray}

\subsection{Correlation functions and difference equations}

Let us consider the $2N$-point correlation function:
\begin{eqnarray}
&&G^{(i)}(z_1,\cdots,z_N|z_{N+1},\cdots,z_{2N})
\nonumber\\
&=&
\sum_{j_1=0}^{n-1}\cdots \sum_{j_N=0}^{n-1}
\sum_{j_{N+1}=0}^{n-1}\cdots\sum_{j_{2N}=0}^{n-1}
v_{j_1}^* \otimes \cdots \otimes v_{j_N}^* \otimes
v_{j_{N+1}} \otimes \cdots \otimes v_{j_{2N}}\\
&\times&
G^{(i)}(z_1,\cdots,z_N|z_{N+1},\cdots,z_{2N})^
{j_1\cdots j_N}_{j_{N+1} \cdots j_{2N}}, \nonumber
\end{eqnarray}
where 
\begin{eqnarray}
&&G^{(i)}(z_1,\cdots,z_N|z_{N+1},\cdots,z_{2N})^
{j_1\cdots j_N}_{j_{N+1} \cdots j_{2N}}\nonumber\\
&=&~_B\langle i |
\Phi^{*(i,i-1)}_{j_1}(z_1)
\cdots \Phi^{*(i-N+1,i-N)}_{j_N}(z_N
)\Phi^{(i-N,i-N+1)}_{j_{N+1}}(z_{N+1})
\cdots \Phi^{(i-1,i)}_{j_{2N}}(z_{2N}
)|i\rangle_B.
\end{eqnarray}
In order to derive $q$-difference equations,
we use the commutation relations of
vertex operators and the action formulae of vertex
operators to the boundary state.
In what follows we assume that 
$K^{*(i)}(z)$ is a diagonal matrix 
whose diagonal elements are given by
(\ref{C1}), (\ref{C2}) and (\ref{C3}).

The commutation relations between vertex operators 
of different types are given as follows \cite{DO}: 
\begin{eqnarray}
\Phi^{(i,i+1)}_{j}(z_2)
\Phi^{*(i+1,i)}_{j}(z_1)&=&
\sum_{k=0}^{n-1}
R^{(i) V^* V }(z_1/z_2)_{j,j}^{k,k}
\Phi^{*(i,i-1)}_{k}(z_1)\Phi^{(i-1,i)}_{k}(z_2),
\label{commuteV*V}\\
\Phi^{(i,i+1)}_{k}(z_2)
\Phi^{*(i+1,i)}_{j}(z_1)&=&
r^{(i) V^* V }(z_1/z_2)
\Phi^{*(i,i-1)}_{j}(z_1)\Phi^{(i-1,i)}_{k}(z_2),~(j\neq k),
\end{eqnarray}
and
\begin{eqnarray}
\Phi^{*(i,i-1)}_{j}(z_2)
\Phi^{(i-1,i)}_{j}(z_1)&=&
\sum_{k=0}^{n-1}R^{(i) V V^* }(z_1/z_2)_{j,j}^{k,k}
\Phi^{(i,i+1)}_{k}(z_1)\Phi^{*(i+1,i)}_{k}(z_2),
\label{commuteVV*}\\
\Phi^{*(i,i-1)}_{k}(z_2)
\Phi^{(i-1,i)}_{j}(z_1)&=&
r^{(i) V V^* }(z_1/z_2)
\Phi^{(i,i+1)}_{j}(z_1)\Phi^{*(i+1,i)}_{k}(z_2),~~(j\neq k).
\end{eqnarray}
Here the nonzero components are 
\begin{eqnarray}
R^{(i) V^* V }(z)_{j,j}^{k,k}&=&
r^{(i)V^*V}(z)\times\left\{\begin{array}{cc}
b(z),&~~j=k,\\
c(z),&~~j>k,\\
zc(z),&~~j<k,
\end{array}\right.
\label{RV*V}
\end{eqnarray}
and
\begin{eqnarray}
R^{(i) V V^* }(z)_{j,j}^{k,k}&=&
r^{(i)VV^*}(z)
\times\left\{\begin{array}{cc}
b(q^{2n}z),&~~j=k,\\
q^{2n}zc(q^{2n}z)q^{2(k-j)},&~~j>k,\\
c(q^{2n}z)q^{2(k-j)},&~~j<k,
\end{array}\right.\label{RVV*}
\end{eqnarray}
where 
\begin{eqnarray}
r^{(i)V^*V}(z)=-qz^{-\delta_{i,0}}
\frac{(z;q^{2n})_\infty (q^{2n+2}z^{-1};q^{2n})_\infty}
{(q^2z;q^{2n})_\infty (q^{2n}z^{-1};q^{2n})_\infty},\\
r^{(i)VV^*}(z)=-q^{-1}z^{-\delta_{i,0}}
\frac{(q^{2n}z;q^{2n})_\infty (q^{2}z^{-1};q^{2n})_\infty}
{(q^{2n+2}z;q^{2n})_\infty (z^{-1};q^{2n})_\infty}.
\end{eqnarray}
The commutation relations between the dual vertex operators
are given as 
\begin{eqnarray}
\Phi^{*(i+2,i+1)}_{j_2}(z_2)
\Phi^{*(i+1,i)}_{j_1}(z_1)=
\sum_{k_1,k_2=0
\atop{k_1+k_2=j_1+j_2}}^{n-1}
R^{(i) V^* V^* }(z_1/z_2)_{j_1,j_2}^{k_1,k_2}
\Phi^{*(i+2,i+1)}_{k_1}(z_1)\Phi^{*(i+1,i)}_{k_2}(z_2).
\label{commuteV*V*}\end{eqnarray}
Here the nonzero components are
\begin{eqnarray}
R^{(i) V^* V^* }(z_1/z_2)_{j_1,j_2}^{k_1,k_2}&=&
r^{(i)V^*V^*}(z)\times\left\{\begin{array}{cc}
1,&~~j_1=j_2=k_1=k_2,\\
b(q^2z),&~~j_1=k_1 \neq j_2=k_2,\\
-qz c(q^2),&~~j_1=k_2<j_2=k_1,\\
-q c(q^2),&~~j_1=k_2>j_2=k_1,
\end{array}\right.\label{RV*V*}
\end{eqnarray}
where 
\begin{eqnarray}
r^{(i)V^*V^*}(z)=r^{(i)VV}(z).
\end{eqnarray}

Now we are in a position to derive boundary quantum 
Knizhnik-Zamolodchikov equations, 
which is a version of Cherednik's
equation \cite{Che}. From the commutation relations 
(\ref{commuteVV}), (\ref{commuteV*V}), 
(\ref{commuteVV*}), (\ref{commuteV*V*}) and 
the boundary state ideitities (\ref{eq:dual-KP}) 
we obtain the following $q$-difference equations: 
\begin{eqnarray}
&&G^{(i)}(z_1 \cdots q^{-2n}z_j \cdots z_N|
z_{N+1} \cdots z_{2N})\nonumber \\
&=&
R_{j j-1}^{V^*V^*}(z_j/(q^{2n}z_{j-1}))
\cdots R_{j 1}^{V^*V^*}(z_j/(q^{2n}z_{1}))K_j^{(i)}
(z_j/q^{2n})\nonumber
\\
&\times&
R_{1 j}^{V^*V^*}(z_1 z_j/q^{2n}) \cdots
R_{j-1 j}^{V^* V^*}(z_{j-1} z_j/q^{2n})
R_{j+1 j}^{V^*V^*}(z_{j+1} z_j/q^{2n}) \cdots
R_{N j}^{V^* V^*}(z_{N} z_j/q^{2n})
\label{BQKZ1}\\
&\times&
R_{N+1 j}^{V V^*}(z_{N+1}z_j/q^{2n})\cdots
R_{2N j}^{V V^*}(z_{2N}z_j/q^{2n}) K_j^{*(i)}
(q^{n}/z_j)R_{j 2N}^{V^* V}(z_j/z_{2N}) \cdots
R_{j N}^{V^* V}(z_j/z_N)
\nonumber\\
&\times&
R_{j N+1}^{V^* V^*}(z_j/z_{N+1})\cdots
R_{j j+1}^{V^* V^*}(z_j/z_{j+1})
G^{(i)}(z_1 \cdots z_N|z_{N+1}\cdots z_{2N}),\nonumber
\end{eqnarray}
and
\begin{eqnarray}
&&G^{(i)}(z_1\cdots z_N|z_{N+1}\cdots q^{-2n}z_j \cdots z_{2N})
\nonumber \\
&=&
R_{jj-1}^{VV}(z_j/(q^{2n}z_{j-1}))
\cdots R_{jN+1}^{VV}(z_j/(q^{2n}z_{N+1}))
R_{jN}^{VV^*}(z_j/(q^{2n}z_N))
\cdots R_{j1}^{VV^*}(z_j/(q^{2n}z_1))
\nonumber \\
&\times&K^{*(i)}_j(q^n/z_j) 
R_{1j}^{V^*V}(z_1z_j)\cdots
R_{Nj}^{V^*V}(z_Nz_j)
R_{N+1j}^{VV}(z_{N+1}z_j)\cdots
R_{j-1j}^{VV}(z_{j-1}z_j)\label{BQKZ2}\\
&\times&
R_{j+1j}^{VV}(z_{j+1}z_j)\cdots
R_{2Nj}^{VV}(z_{2N}z_j)K_j^{(i)}(z_j)\nonumber\\
&\times&
R_{j2N}^{VV}(z_j/z_{2N})\cdots
R_{jj+1}^{VV}(z_j/z_{j+1})
G^{(i)}(z_1\cdots z_N|z_{N+1}\cdots z_{2N}).\nonumber
\end{eqnarray}
Here the coefficient matrces are given by
(\ref{RVV}),(\ref{K}),(\ref{C1}),
(\ref{C2}),(\ref{C3}),(\ref{RV*V}),
(\ref{RVV*}) and (\ref{RV*V*}).

For $N=1$, the equations (\ref{BQKZ1}) and (\ref{BQKZ2})
are as follows:
\begin{eqnarray}
G^{(i)}(q^{-2n}z_1|z_2)&=&
K_1^{(i)}(z_1/q^{2n})R_{21}^{VV^*}(z_2z_1/q^{2n})K_1^{*(i)}
(q^{n}/z_1)
R_{12}^{V^*V}(z_1/z_2)G^{(i)}(z_1|z_2),\\
G^{(i)}(z_1|q^{-2n}z_2)&=&
R_{21}^{VV^*}(z_2/(q^{2n}z_1))K_2^{*(i)}(q^n/z_2)
R_{12}^{V^*V}(z_1z_2)K_2^{(i)}(z_2)
G^{(i)}(z_1|z_2).
\end{eqnarray}

\section{Two point functions}
The purpose of this section
is to perform explicit calculations of two point functions
for free boundary condition.
In what follows we consider the case 
$i=L=M=0$ and $N=1$.
In this case the boundary K-matrices $K^{(0)}(z)$ and
$K^{*(0)}(z)$ become scalar matrices, i.e.
$$
K^{(0)}(z)=\frac{\varphi^{(0)}(z)}
{\varphi^{(0)}(z^{-1})}\times {\rm id},~~~
K^{*(0)}(z)=\frac{\varphi^{*(0)}(z)}
{\varphi^{*(0)}(z^{-1})} \times {\rm id}.
$$
The boundary quantum 
Knizhnik-Zamolodchikov equations thus reduces to:

\begin{eqnarray}
G^{(0)}(q^{-2n}z_1|z_2)&=&
\frac{\varphi^{(0)}(z_1/q^{2n})}
{\varphi^{(0)}(q^{2n}/z_1)}
\frac{\varphi^{*(0)}(q^n/z_1)}
{\varphi^{*(0)}(z_1/q^{n})}
R_{21}^{VV^*}(z_2z_1/q^{2n})
R_{12}^{V^*V}(z_1/z_2)G^{(0)}(z_1|z_2),\\
G^{(0)}(z_1|q^{-2n}z_2)&=&
\frac{\varphi^{(0)}(z_2)}
{\varphi^{(0)}(1/z_2)}
\frac{\varphi^{*(0)}(q^n/z_2)}
{\varphi^{*(0)}(z_2/q^n)}
R_{21}^{VV^*}(z_2/(q^{2n}z_1))
R_{12}^{V^*V}(z_1z_2)
G^{(0)}(z_1|z_2).
\end{eqnarray}

Let us now introduce the scalar function $r(z_1|z_2)$ by
\begin{eqnarray}
r(z_1|z_2)=A(z_1)A(q^nz_2)B(z_1z_2)B(z_1/z_2),
\end{eqnarray}
where 
\begin{eqnarray}
A(z)&=&\frac{(q^{2n+2}z^2;q^{2n},q^{4n})_\infty
(q^{4n+2}/z^{2};q^{2n},q^{4n})_\infty}
{(q^{4n}z^2;q^{2n},q^{4n})_\infty 
(q^{6n}/z^2;q^{2n},q^{4n})_\infty},\\
B(z)&=&\frac{
(q^{2n}z;q^{2n},q^{2n})_\infty
(q^{2n}/z;q^{2n},q^{2n})_\infty
}{
(q^{2n+2}z;q^{2n},q^{2n})_\infty
(q^{2n+2}/z;q^{2n},q^{2n})_\infty
}.
\end{eqnarray}
Note that the function $r(z_1|z_2)$ satisfies
\begin{eqnarray}
r(q^{-2n}z_1|z_2)=q^{-2n}z_1^2
r^{VV^*}(z_1z_2/q^{2n})r^{V^*V}(z_1/z_2)
\frac{\varphi^{(0)}(z_1/q^{2n})}
{\varphi^{(0)}(q^{2n}/z_1)}
\frac{\varphi^{*(0)}(q^n/z_1)}
{\varphi^{*(0)}(z_1/q^{n})}\times
r(z_1|z_2),
\\
r(z_1|q^{-2n}z_2)=q^{-2n}z_2^2
r^{VV^*}(z_2/(q^{2n}z_1))r^{V^*V}(z_1z_2)
\frac{\varphi^{(0)}(z_2)}
{\varphi^{(0)}(1/z_2)}
\frac{\varphi^{*(0)}(q^n/z_2)}
{\varphi^{*(0)}(z_2/q^n)}\times
r(z_1|z_2).
\end{eqnarray}
Let $\bar{G}(z_1|z_2)_j$ be the auxiliary 
function defined by
\begin{eqnarray}
\bar{G}(z_1|z_2)_j={r(z_1|z_2)}^{-1}G^{(0)}(z_1|z_2)_j^j.
\end{eqnarray}
Then we have
\begin{eqnarray}
\sum_{j=0}^{n-1}
\bar{G}(q^{-2n}z_1|z_2)_j&=&
\frac{1-q^{2n}/(z_1z_2)}{1-z_1z_2}
\frac{1-q^{2n}z_2/z_1}{1-z_1/z_2}
\sum_{j=0}^{n-1}~
\bar{G}(z_1|z_2)_j,\\
\sum_{j=0}^{n-1}
\bar{G}(z_1|q^{-2n}z_2)_j&=&
\frac{1-q^{2n}/(z_1z_2)}{1-z_1z_2}
\frac{1-q^{2n}z_1/z_2}{1-z_2/z_1}
\sum_{j=0}^{n-1}~
\bar{G}(z_1|z_2)_j.
\end{eqnarray}
From these we obtain 
\begin{eqnarray}
&&\sum_{j=0}^{n-1}~_B\langle 0|
\Phi_j^{*(0,1)}(z_1)
\Phi_j^{(1,0)}(z_2)
|0\rangle_B
=c_0
~r(z_1|z_2)\times \\
&\times&
\left\{(q^{2n}z_1/z_2;q^{2n})_\infty
(q^{2n}z_2/z_1;q^{2n})_\infty
(q^{2n}z_1z_2;q^{2n})_\infty
(q^{2n}/(z_1z_2);q^{2n})_\infty
\right\}^{-1},\nonumber
\end{eqnarray}
where $c_0$ is a constant independent of spectral
parameters $z_1, z_2$.
By specializing the spectral parameters $z_1=z_2$,
we have
\begin{eqnarray}
c_0=g_n{}^{-1}\times~_B\langle 0 | 0\rangle_B \times
\left\{\frac{(q^{2n+2};q^{2n},q^{2n})_\infty}
{(q^{4n};q^{2n},q^{2n})_\infty}\right\}^2,
\end{eqnarray}
where the norm ${}_B\langle 0 | 0\rangle_B$ is given 
as follows \cite{FK}
\begin{eqnarray}
~_B\langle 0 | 0\rangle_B
=
\frac{1}{\sqrt{(q^{4n};q^{4n})_\infty}}
\prod_{j=1}^{n-1}\left\{
\frac{\sqrt{(q^{4n+2-2j};q^{4n})_\infty 
(q^{4n-2-2j};q^{4n})_\infty}}{(q^{4n-2j};q^{4n})_\infty}
\right\}^{j(n-j)}.\nonumber
\end{eqnarray}

Let $\omega$ satisfy $\omega^n=1$ and 
$\omega\neq 1$. Then we have
\begin{eqnarray}
\sum_{j=0}^{n-1}
(q^2 \omega)^j \bar{G}(q^{-2n}z_1|z_2)_j&=&
q^{2n}z_1^{-2}\sum_{j=0}^{n-1}
(q^2 \omega)^j
\bar{G}(z_1|z_2)_j,\\
\sum_{j=0}^{n-1}
(q^2 \omega)^j \bar{G}(z_1|q^{-2n}z_2)_j&=&
q^{2n}z_2^{-2}\sum_{j=0}^{n-1}
(q^2 \omega)^j
\bar{G}(z_1|z_2)_j.
\end{eqnarray}
From these we obtain 
\begin{eqnarray}
&&\sum_{j=0}^{n-1}
(q^2 \omega^k)^j 
~_B\langle 0|
\Phi_j^{*(0,1)}(z_1)
\Phi_j^{(1,0)}(z_2)
|0\rangle_B\nonumber\\
&=&c_k 
~r(z_1|z_2)
\times
\{(-q^{2n}z_1^2;q^{4n})_\infty 
(-q^{2n}/z_1^2;q^{4n})_\infty
(-q^{2n}z_2^2;q^{4n})_\infty 
(-q^{2n}/z_2^2;q^{4n})_\infty
\}^{-1}.
\end{eqnarray}
Here
$c_k$ are constants independent of
spectral parameters $z_1, z_2$.

~

{\it Acknoeledgements.}~~
We wish to thank Prof. A. Kuniba for his interest
to this work.
TK was partly supported by Grant-in-Aid for
Encouragements for Young Scientists (A) from Japan Society
for the Promotion of Science. (11740099)

\appendix

\section{Bosonization of vertex operators 
in $U_q (\widehat{sl_n})$}
For readers' convenience, we summarize 
the results of bosonizations
of the vertex operators \cite{Koy}.

Let ${\mathbb C}[\bar{P}]$
be the ${\mathbb C}$-algebra generated by the symbols
$\{e^{\alpha_2},\cdots,e^{\alpha_{n-1}},
e^{\bar{\Lambda}_{n-1}}\}$
which satisfy the following defining relations:
\begin{eqnarray*}
e^{\alpha_i}e^{\alpha_j}=(-1)^{(\alpha_i|\alpha_j)}
e^{\alpha_j}e^{\alpha_i},~~(2\leqq i,j \leqq n-1),\\
e^{\alpha_i}e^{\bar{\Lambda}_{n-1}}=
(-1)^{\delta_{i,n-1}}e^{\bar{\Lambda}_{n-1}}
e^{\alpha_i},~~(2\leqq i \leqq n-1).
\end{eqnarray*}
For $\alpha=m_2 \alpha_2+\cdots +m_{n-1}\alpha_{n-1}
+m_n \bar{\Lambda}_{n-1}$, we denote
$e^{m_2 \alpha_2}\cdots e^{m_{n-1}\alpha_{n-1}}
e^{m_n \bar{\Lambda}_{n-1}}$
by $e^{\alpha}$.
Let $((\alpha_s|\alpha_t))_{1\leqq s,t \leqq n-1}$ 
stand for the A-type Catran matrix whose matrix 
element $(\alpha_s|\alpha_t)$ is an integer. 
Let ${\mathbb C}[\bar{Q}]$ be the 
${\mathbb C}$-subalgebra of ${\mathbb{C}}[\bar{P}]$
generated by the symbols
$\{e^{\alpha_1}, \cdots, e^{\alpha_{n-1}}\}$
which satisfy the following defining relations:
\begin{eqnarray*}
e^{\alpha_i}e^{\alpha_j}=
(-1)^{(\alpha_i|\alpha_j)}e^{\alpha_j}
e^{\alpha_i},
~~(1\leqq i,j \leqq n-1).
\end{eqnarray*}
Note that
\begin{eqnarray*}
\alpha_1=-\sum_{r=2}^{n-1}r \alpha_r +n \bar{\Lambda}_{n-1},
~~
\bar{\Lambda}_i=-
\sum_{r=i+1}^{n-1}(r-i)\alpha_r
+(n-i)\bar{\Lambda}_{n-1}.
\end{eqnarray*}

Let us consider the ${\mathbb C}$-algebra
generated by the bosons 
$a_s(k)$ $(s\in \{1,\cdots, n-1 \}, k \in {\mathbb Z})$ 
which satisfy the following defining relations:
$$
[a_s(k),a_t(l)]=\delta_{k+l,0}\frac{
[(\alpha_s|\alpha_t)k][k]}{k}.
$$
The highset weight module $V(\Lambda_i)$ is realized as
$$
V(\Lambda_i)={\mathbb C}[a_s(-k),
~(s\in \{1,\cdots,n-1\},k \in {\mathbb Z}\geqq 0)]
\otimes {\mathbb C}[\bar{Q}]e^{\bar{\Lambda}_i}.
$$
We consider ${\mathbb{C}}[\bar{Q}]e^{\bar{\Lambda}_i}$
as a subspace of
${\mathbb{C}}[\bar{P}]$.
Here the actions of the operators
$a_s(k), \partial_{\alpha}, e^{\alpha}$
on $V(\Lambda_i)$
are defined as follows:
\begin{eqnarray*}
a_s(k) f \otimes e^{\beta}=
\left\{
\begin{array}{cc}
a_s(k) f \otimes e^{\beta}, &~(k<0),\\
\left[a_s(k),f\right] \otimes e^{\beta},&~(k>0),
\end{array}\right.
\end{eqnarray*}
\begin{eqnarray*}
\partial_{\alpha} f \otimes e^{\beta}&=&(\alpha|\beta)
f \otimes e^{\beta}.\\
e^{\alpha} f \otimes e^{\beta}&=&f\otimes e^{\alpha}
e^{\beta}.
\end{eqnarray*}
The inner product is explicitly given as follows: 
\begin{eqnarray*}
(\alpha_i|\bar{\Lambda}_j)=\delta_{i,j},~~
(\bar{\Lambda}_i|\bar{\Lambda}_j)=\frac{i(n-j)}{n},
~~(1\leqq i\leqq j \leqq n-1).
\end{eqnarray*}
\begin{eqnarray*}
\Phi_{n-1}^{(i,i+1)} (z) &=& 
e^{P(z)}e^{Q(z)}e^{\bar{\Lambda}_{n-1}}
(q^{n+1}z)^{\partial_{\bar{\Lambda}_{n-1}}+\frac{n-i-1}{n}}
(-1)^{(\partial_{\bar{\Lambda}_1}-\frac{n-i-1}{n})(n-1)+\frac{1}{2}(n-i)(n-i-1)},\\
\Phi_0^{* (i+1,i)} (z) &=& e^{P^*(z)}e^{Q^*(z)}e^{\bar{\Lambda}_1}
((-1)^{n-1}qz)^{\partial_{\bar{\Lambda}_1}+\frac{i}{n}}q^i
(-1)^{in+\frac{1}{2}i(i+1)},
\end{eqnarray*}
\begin{eqnarray}
\Phi_j^{(i-1 i)}(z)&=&
c_j\oint \cdots \oint_{C_j} \frac{dw_{j+1}}{2\pi i w_{j+1}}
\cdots \frac{dw_{n-1}}{2\pi i w_{n-1}}
\frac{w_{j+1}}{z}
\frac{1}{(1-qw_{n-1}/z)(1-qz/w_{n-1})}
\nonumber\\
&\times&
\frac{1}{(1-qw_{n-1}/w_{n-2})(1-qw_{n-2}/w_{n-1})
\cdots (1-qw_{j+2}/w_{j+1})(1-qw_{j+1}/w_{j+2})}
\nonumber\\
&\times&
:\Phi_{n-1}^{(i-1 i)}(z)
X_{n-1}^-(q^{n+1}w_{n-1})\cdots
X_{j+1}^-(q^{n+1}w_{j+1}):,
\end{eqnarray}
and
\begin{eqnarray}
\Phi_j^{*(i i+1)}(z)
&=&c_j^*
\oint \cdots \oint_{C_j^*} \frac{dw_1}{2\pi i w_1}
\cdots \frac{dw_{j}}{2\pi i w_j}
\frac{w_j}{z}\frac{1}
{\left(1-qz/w_1\right)\left(1-qw_1/z\right)}
\nonumber \\
&\times&
\frac{1}{(1-qw_1/w_2)(1-qw_2/w_1)\cdots 
(1-qw_{j-1}/w_j)(1-qw_j/w_{j-1})}\nonumber\\
&\times&
:\Phi_0^{*(i i+1)}(z)
X_1^-(qw_1)\cdots X_j^-(qw_j):,
\end{eqnarray}
where $c_j, c_j^*$ are appropriate constants.
The contours $C_j, C_j^*$ encircle $w_l=0$ 
anti-clockwise in such a way that
\begin{eqnarray*}
C_j:&& |q|< | w_{n-1}/z |<|q^{-1}|,~~
|q|< | w_{l}/w_{l+1} |<|q^{-1}|,~(l=j+1,\cdots, n-2),\\
C_j^*:&& |q|< | w_{1}/z |<|q^{-1}|,~~
|q|< | w_{l+1}/w_{l} |<|q^{-1}|,~(l=1,\cdots, j-1). \\
\end{eqnarray*}
Here we have used 
\begin{eqnarray*}
X_j^-(w)=e^{R_j^-(w)}e^{S_j^-(w)}e^{-\alpha_j}
w^{-\partial_{\alpha_j}},
\end{eqnarray*}
\begin{eqnarray*}
P(z) = \sum_{k=1}^{\infty} a_{n-1}^*(-k)q^{\frac{2n+3}{2}k}z^k, &&
Q(z) = \sum_{k=1}^{\infty} a_{n-1}^*(k)q^{-\frac{2n+1}{2}k}z^{-k},\\
P^*(z) = \sum_{k=1}^{\infty} a_1^*(-k)q^{\frac{3}{2}k}z^k, &&
Q^*(z) = \sum_{k=1}^{\infty} a_1^*(k)q^{-\frac{1}{2}k}z^{-k},\\
R_j^{-}(w) = -\sum_{k=1}^{\infty}
\frac{a_j(-k)}{[k]}q^{\frac{k}{2}}w^k,&&
S_j^{-}(w) = \sum_{k=1}^{\infty}
\frac{a_j(k)}{[k]}q^{\frac{k}{2}}w^{-k},\\
a_{n-1}^*(k) = \sum_{l=1}^{n-1}\frac{-[lk]}{[k][nk]}a_l(k), &&
a_1^*(k) = \sum_{l=1}^{n-1}\frac{-[(n-l)k]}{[k][nk]}a_l(k).
\end{eqnarray*}
\begin{eqnarray*}
[a_j(k),a_{n-1}^*(-k)]=\delta_{j,n-1}\frac{[k]}{k},~~
[a_j(k),a_1^*(-k)]=\delta_{j,1}\frac{[k]}{k}.
\end{eqnarray*}

\section{Bosonization of the boundary vacuum states}
For readers' convenience we summarize the bosonic 
formulae of the boundary vacuum \cite{FK}.
Let us set the symmetric matrix as
\begin{eqnarray}
\hat{I}_{s,t}(k)=\left\{
\begin{array}{ll}
0,& (st=0),\\
\frac{[sk][(n-t)k]}{[k]^2[nk]},& 
(1\leqq s \leqq t \leqq n-1),\\
\frac{[tk][(n-s)k]}{[k]^2[nk]},& 
(1\leqq t \leqq s \leqq n-1).
\end{array}
\right.
\end{eqnarray}
Let us consider the $\mathbb{C}$-algebra generated by 
the bosons $a_s(k)$ $
(s \in \{1,\cdots,n-1\}, k \in {\mathbb{Z}})$
which satisfy the following defining relations:
\begin{eqnarray*}
[a_s(k),a_t(l)]=\delta_{k+l,0}\frac{
[(\alpha_s|\alpha_t)k][k]}{k}, 
\end{eqnarray*}
where I$(\alpha_s|\alpha_t)$ is an element
of A-type Cartan matrix. 

The boundary state 
has the form 
\begin{eqnarray*}
|i\rangle_B=e^{F_i}|i\rangle , ~~~~
F_i=\sum_{s,t=1}^{n-1}\sum_{k=1}^\infty
\alpha_{s,t}(k)
a_s(-k)a_t(-k)+\sum_{s=1}^{n-1}\sum_{k=1}^\infty
\beta_s^{(i)}(k)a_s(-k). 
\end{eqnarray*}
Here the coefficients of the quadratic part are given by
\begin{eqnarray}
\alpha_{s,t}(k)=
\frac{-kq^{2(n+1)k}}{2[k]}\times \hat{I}_{s,t}(k), 
\end{eqnarray}
and those of the linear part are given by
\begin{eqnarray}
\beta_j^{(i)}(k)&=&(q^{(n+3/2)k}-q^{(n+1/2)k})\theta_k
\sum_{s=1}^{n-1}\hat{I}_{j,s}(k)\\
&+&\left\{
\begin{array}{ll}
0,& (C1),\\
\hat{I}_{j,L}(k)q^{(2n-2M+L+1/2)k}r^k-\hat{I}_{j,M}(k)
q^{(2n-M+1/2)k}r^k,& (C2),\\
-\hat{I}_{j,L}(k)q^{(2M-L+1/2)k}r^{-k}
+\hat{I}_{j,M}(k)q^{(M+1/2)k}r^{-k},& (C3), 
\end{array}\right.
\end{eqnarray}
where
\begin{eqnarray*}
\theta_k=\left\{\begin{array}{cc}
0,& \mbox{$k$ is odd},\\
1,& \mbox{$k$ is even}.
\end{array}\right.
\end{eqnarray*}

The dual boundary state has the form 
\begin{eqnarray}
~_B\langle i |=\langle i |e^{G_i}, ~~~~
G_i=\sum_{s,t=1}^{n-1}\sum_{k=1}^\infty\gamma_{s,t}(k)
a_s(k)a_t(k)+\sum_{s=1}^{n-1}\sum_{k=1}^\infty
\delta_s^{(i)}(k)
a_s(k). 
\end{eqnarray}
Here the coefficients of the quadratic part are given by
\begin{eqnarray}
\gamma_{s,t}(k)=\frac{-kq^{-2k}}{2[k]}\times \hat{I}_{s,t}(k),
\end{eqnarray}
and those of the linear part are given by
\begin{eqnarray}
\delta_{j}^{(i)}(k)&=&-(q^{-k/2}-q^{-3k/2})\theta_k
\sum_{s=1}^{n-1}\hat{I}_{j,s}(k)\\
&+&
\left\{
\begin{array}{ll}
0,& (C1),\\
q^{(L-3/2)k}r^k
\hat{I}_{j,L}(k)-q^{(2L-M-3/2)k}r^k
\hat{I}_{j,M}(k),& (C2),\\
-q^{(-L-3/2)k}r^{-k}\hat{I}_{j,L}(k)
+q^{(M-2L-3/2)k}r^{-k}\hat{I}_{j,M}(k),&(C3).
\end{array}\right.
\end{eqnarray}


\begin{thebibliography}{99}
\bibitem{FR}Frenkel I B and Reshetikhin N Y:
Quantum affine algebras and holonomic difference equations,
{\it Commun. Math. Phys.} {\bf 146}, 1-60, (1992).
\bibitem{Sm1} Smirnov, F A: Dynamical 
Symmetries of Massive Integrable Models 1:
Int. J. Mod. Phys. {\bf 7A, Suppl. 1B} (1992) 813-837;
2: ibid. 839-858.
\bibitem{JM} Jimbo M and Miwa T: 1994, 
{\it Algebraic analysis of solvable lattice models, 
CBMS Regional Conferences Series in Mathematics.} 
\bibitem{S}Smirnov F A:
{\it Form Factors in Completely Integrable Models of
Quantum Field Theory},
World Scientific, Singapore, 1992.
\bibitem{JMN}Jimbo M, Miwa T and Nakayashiki A:
Difference equations for the correlation functions of
the eight-vertex model, {\it J. Phys.}{\bf A26},
2199-2209, (1993), 
\bibitem{Skl}Sklyanin E K:
Boundary conditions for integrable quantum systems,
{\it J. Phys.}{\bf A21}, 2375-2389, (1988).
\bibitem{GZ}Ghoshal S and Zamolodchikov A:
Boundary S-matrix and boundary state in two
dimensianal integrable quantum field theory,
{\it Int. J. Mod. Phys.} {\bf A9}, 3841-3886 ;
Erratum ibid. 4353, (1994).
\bibitem{JKKKM}Jimbo M, Kedem R, Kojima T, Konno H and
Miwa T:
XXZ chain with a boundary, {\it Nucl. Phys.}{\bf B [FS]},
437-470, (1995).
\bibitem{MW}Miwa T and Weston R:
Boundary ABF Models, {\it Nucl. Phys.}{\bf B [PM]},
517-545, (1997).
\bibitem{HSWY}Hou B Y, Shi K J, Wang Y S and Yang W L:
Bosonization of quantum sine-Gordon field
with boundary,
{\it Int. J. Mod. Phys.}{\bf A12}, 1711-1741, (1997).
\bibitem{FK}Furutsu H and Kojima T:
$U_q(\widehat{sl_n})$-analog of the XXZ chain with a boundary,
solv-int/9905009.
\bibitem{FKQ}Furutsu H, Kojima T and Quano Y.-H: 
Form factors of the $SU(2)$ invariant massive Thirring
model with boundary reflection,
to appear in {\it Int. J. Mod. Phys.}{\bf A} (2000).
\bibitem{H}Hara Y:
Correlation functions of the XYZ model with a boundary,
[math-ph/9910046]
\bibitem{JKKMW}Jimbo M, Kedem R, Konno K, Miwa T and
Weston R: Diffence Equations in Spin Chains with a Boundary,
{\it Nucl. Phys.} {\bf B448[FS]}, 429-456, (1995).
\bibitem{Koy}Koyama Y: Staggered Polarization of Vertex Models
with $U_q(\widehat{sl_n})$-Symmetry,
{\it Commun. Math. Phys.} {\bf 164}, 277-291, (1994).
\bibitem{DO}Date E and Okado M: Caluculation of excited 
spectra of the spin model related with the vector representation
of the quantized affine algebra of type
$A_n^{(1)}$,
{\it Int. J. Mod. Phys.}{\bf 9}, 399-417, (1994).
\bibitem{Che}Cherednik I V: Factorizing particles on a half-line
and root systems,
{\it Theor. Math. Phys.}{\bf 61},
977-983, (1984).

\end{thebibliography}
\end{document}